\documentclass[%
mathleft,%
]{an}
\usepackage{graphicx}
\usepackage{times}
\begin{document}

% The following seven commands are intended for editorial usage and
% should be ignored by the author(s).
\Pagespan{1}{}% Document's page range. 
% If second parameter is left empty, the last page is computed
% automatically.
\Yearpublication{2011}%
\Yearsubmission{2011}%
\Month{1}%   
\Volume{999}%  
\Issue{92}% 
% \DOI{This.is/not.aDOI}% 

\title{The Lockman Hole with LOFAR -- \\
Searching for GPS and CSS sources at low frequencies }

\author{E.\,K. Mahony\inst{1}\fnmsep\thanks{Corresponding author:
  \email{mahony@astron.nl}}
% Example for footnote, note the usage of the \texttt{fnmsep} command
% as separator between institute number and footnote mark}
\and  R. Morganti\inst{1,2}
\and  I. Prandoni\inst{3}
\and  I. van Bemmel\inst{4} on behalf of the LOFAR Surveys Key Science Project
}
\titlerunning{Searching for GPS and CSS sources at low frequencies}
\authorrunning{E.\,K. Mahony et al.}
\institute{
ASTRON, the Netherlands Institute for Radio Astronomy, Postbus 2, 7990 AA, Dwingeloo, The Netherlands
\and 
Kapteyn Astronomical Institute, University of Groningen, Postbus 800, 9700 AV Groningen, The Netherlands
\and 
INAF-Institute of Radioastronomy, Bologna, Italy
\and
Joint Institute for VLBI in Europe, Postbus 2, 7990 AA Dwingeloo, The Netherlands}

\received{XXXX}
\accepted{XXXX}
\publonline{XXXX}


\abstract{%
 The Lockman Hole Project is a wide international collaboration aimed at exploiting the multi-band extensive and deep information available for the Lockman Hole region, with the aim of better characterizing the physical and evolutionary properties of the various source populations detected in deep radio fields. Recent observations with the LOw-Frequency ARray (LOFAR) extends the multi-frequency radio information currently available for the Lockman Hole (from 350 MHz up to 15 GHz) down to 150 MHz, allowing us to explore a new radio spectral window for the faint radio source population. These LOFAR observations allow us to study the population of sources with spectral peaks at lower radio frequencies, providing insight into the evolution of GPS and CSS sources. In this general framework, I present preliminary results from 150\,MHz LOFAR observations of the Lockman Hole field. }

\maketitle

\section{Introduction}

The LOw-Frequency ARray (LOFAR) is a low-frequency radio interferometer based primarily in the Netherlands, with stations spread across Europe. It operates in two frequency bands; the High-Band Antennas (HBA), which cover a frequency range of 110-190\,MHz, and the Low-Band Antennas (LBA) which operate from 10-90\,MHz. There are 38 stations in the Netherlands, 24 of these are located in a densely packed, 2-km wide `core' and the remaining 14 `remote' stations are spread across the Netherlands with baselines up to 120\,km. In addition to the `dutch-array', there are 9 international stations which can be added to the array, extending the maximum baseline out to 1300\,km. The total bandwidth available is 95\,MHz which can be split up into 488 subbands of 0.195\,MHz each. For full details of the capabilities and key science goals of LOFAR we refer the reader to \cite{lofar}.

LOFAR not only allows us to explore the low-frequency radio regime, but the large field of view makes it an ideal survey instrument. At a frequency of 150\,MHz the primary beam size is $\sim 5$\, degrees \footnote{the size of the primary beam varies with the elevation of the target}, meaning that a single 10\,hr pointing results in a sample of thousands of sources. As such, each LOFAR observation can be viewed as a small survey in its own right. 

This combination of deep, low-frequency observations over a wide-field makes LOFAR an ideal telescope for studying the evolution of Gigahertz Peaked Spectrum (GPS) and Compact Steep Spectrum (CSS) sources. Detecting GPS and CSS sources at lower frequencies allows us to better measure the spectral turnover and trace the spectral peak down to lower frequencies. The well known correlation between peak frequency and linear size of the radio source suggests that sources peaking at MHz frequencies would be typical CSS sources, with linear sizes on scales of $\sim$10\,kpc (\cite{odea,snellen,devries}). Studying these sources will therefore allow us to probe the intermediate stages of radio galaxy evolution, bridging the gap beween the young, compact sources detected at high-frequencies and typical FRI/FRIIs. Alternatively, searching for sources peaking at low frequencies has also been suggested as a way to find high redshift GPS sources (\cite{falcke, coppejans}), where the spectral peak is redshifted down to MHz frequencies. In addition, observing these sources at low frequencies allows us to investigate the cause of the spectral turnover. Synchrotron Self Absorption (SSA) is generally the favoured model (\cite{fanti,devries}), but in some cases there is evidence that free-free absoprtion could also play a role (see e.g. \cite{bicknell, tingay}). 

Here, we present recent LOFAR observations of the Lockman Hole field at 150\,MHz. Studying the low-frequency spectral properties of sources in this field can provide insight into the population of GPS and CSS sources we may expect to detect in the upcoming era of low-frequency radio surveys. 

\section{The Lockman Hole Field}

Due to the very low column density of galactic HI, the Lockman Hole is an ideal field for deep observations of extragalactic sources (\cite{lockman}). As such, there is extensive multi-wavelength ancillary data available, including deep optical/NIR data from ground based telescopes (e.g. \cite{Fotopoulou}), midIR/FIR/sub-mm data from the Spitzer and Herschel satellites (\cite{servs,hermes}) and deep X-ray observations from XMM-Newton and Chandra (\cite{xmm,chandra}). However, one of the things that sets the Lockman Hole apart from other famous fields is the wealth of multiwavelength radio data ranging from 350\,MHz - 15\,GHz, essential in searching for GPS and CSS sources. This includes the 15\,GHz 10C survey (\cite{10c, whittam}), deep 1.4\,GHz observations over 7 square degrees observed with the Westerbork Synthesis Radio Telescope (WSRT; \cite{2012rsri.confE..22G}, Prandoni et al., in prep.), 610\,MHz GMRT observations (\cite{garn}) and 345\,MHz WSRT observations (Prandoni et al., in prep.). Recent observations with LOFAR at 150\,MHz and 60\,MHz (Mahony et al., in prep, van Bemmel et al., in prep) extends this multi-frequency information down to much lower frequencies, providing insight into the low-frequency spectral properties of the faint radio source population.

\subsection{LOFAR observations and data reduction}

\begin{figure*}
\hfill\centering{\includegraphics[width=0.7\linewidth, angle=270]{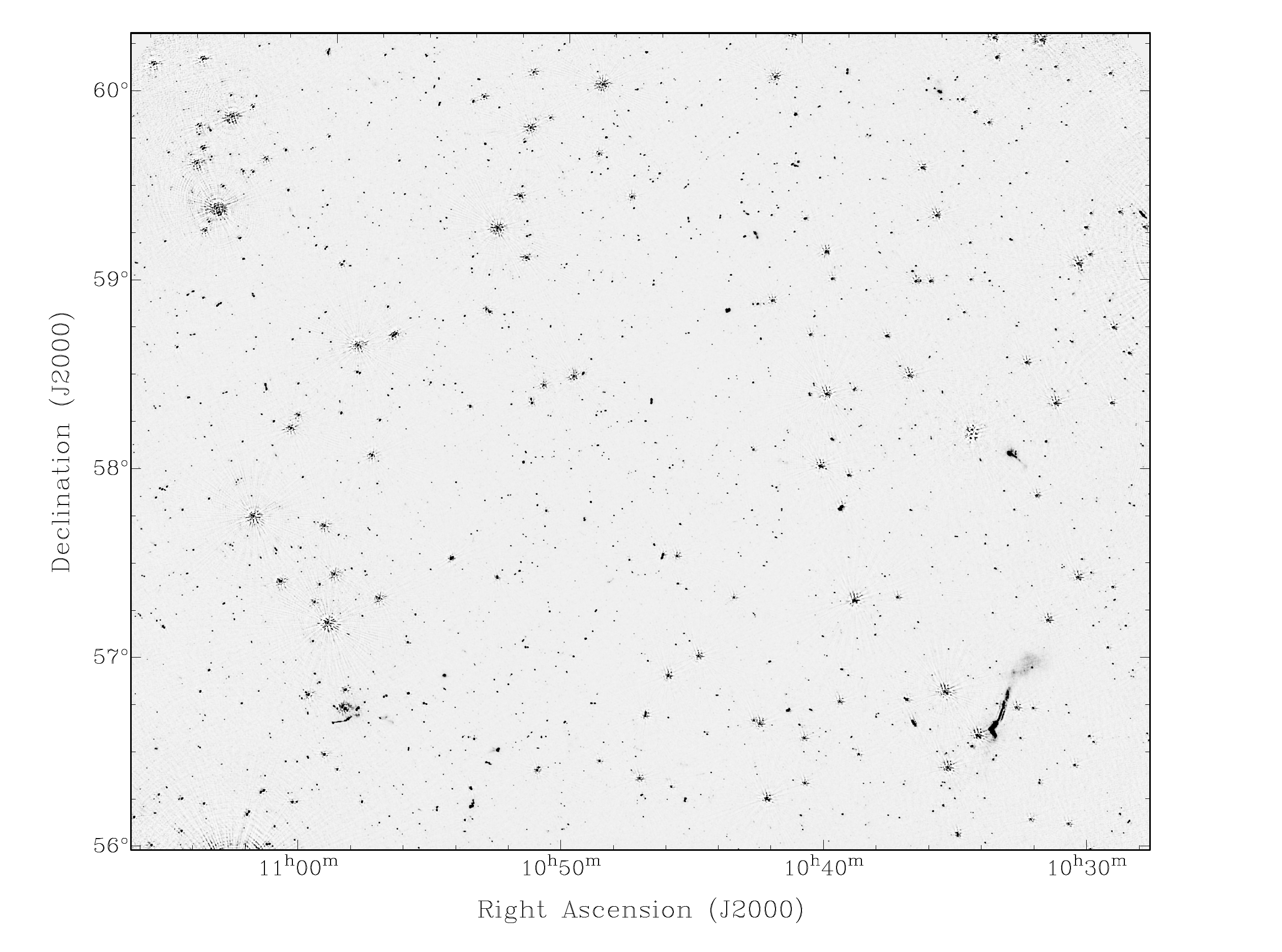} }
%\vspace{1.0cm}
\centering{\includegraphics[width=0.6\linewidth, angle=270]{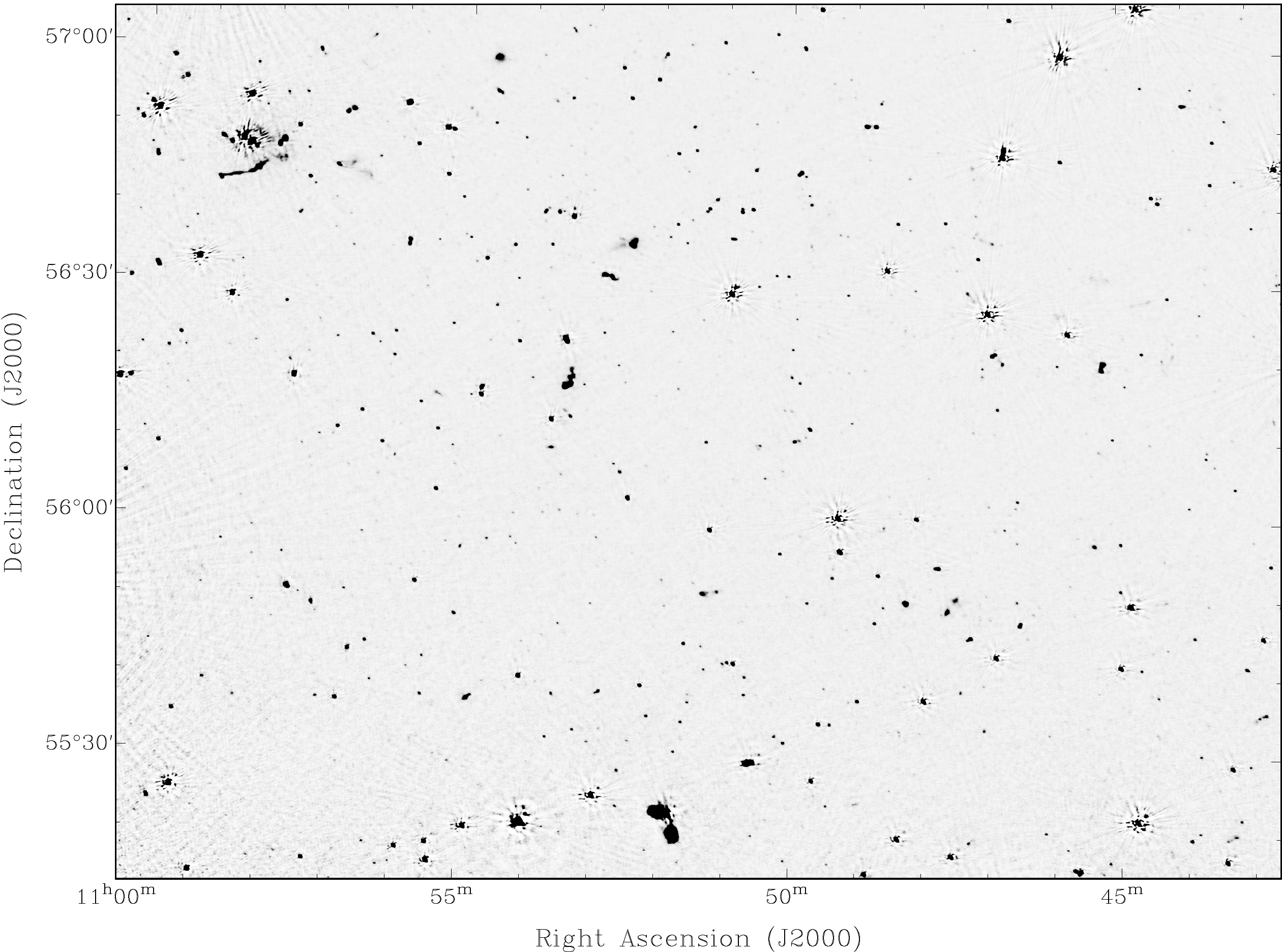}}
\caption{LOFAR images of the Lockman Hole field at 150\,MHz. The top image shows the wide area covered by a single LOFAR pointing and has a resolution of 18.6$\times$14.7\,arcsec, reaching a noise level of 0.15\,mJy/bm in the centre of the field. Over 5000 sources are detected in this image above the 5$\sigma$ level. The bottom image shows a zoomed-in region approximately 1\,degree across. \label{lofarimages}} 
\end{figure*} 

LOFAR observations of the Lockman Hole field (centred at $\alpha$=10h47m, $\delta$=58d05m) were carried out on 18-03-2013 for a total of 10 hrs using the High-Band Antenna (HBA) array. A total of 36 stations (23 core and 13 remote) were used in the observations, corresponding to a range of baseline lengths from $\sim$40\,m -- 120\,kms. A total bandwidth of 72\,MHz was used covering the frequency range 110--182\,MHz. To set the amplitude scale, primary flux calibrators 3C196 and 3C295 were observed for 10 mins before and after the target field. The data were first pre-processed by the standard observatory pipeline (\cite{pipeline}), which included automated RFI flagging using AOflagger (\cite{offringa}) and averaging down to 4 channels and 5\,s integration time per subband. The primary calibrator data was then used to derive the antenna gain solutions for each station per subband using the Black Board Selfcalibration tool (BBS; \cite{bbs}) and the median gain solutions transferred to the target field. The sky models used to calibrate 3C196 and 3C295 were taken from \cite{scaife} so the absolute flux density scale is tied to the RCB flux scale (\cite{rcb}). 

For the phase-calibration, the data was combined into groups of 10 subbands (2\,MHz bandwidth) to ensure there was enough S/N for adequate calibration. Phase-only self-calibration was then carried out using an initial skymodel obtained from LOFAR commissioning observations of the Lockman Hole (\cite{guglielmino}). After the initial phase calibration, a $\sim$35\,Jy source in the field (3C244.1) was peeled to remove ringing artifacts across the field of view. This was followed by another round of phase-only self-calibration. The data were then imaged in 2\,MHz blocks to inspect the image quality across the full bandwidth and significantly noisier images were excluded from further analysis. The remaining 300 subbands were then imaged using the AWimager (\cite{awimager}) with a robust parameter of -0.5.  

The resulting image has a final beam size of 18.6$\times$14.7 arcsec and an rms of $\sim 150 \mu$Jy/bm at the centre of the field. Figure \ref{lofarimages} shows the 150\,MHz LOFAR image of the Lockman Hole field. The top image shows the full LOFAR field of view and the bottom image shows a zoomed in region (approximately 1 degree across). We estimate that due to the combination of errors associated with the calibration (including errors in the beam model and in the absolute flux scale) and the 1-$\sigma$ uncertainties calculated during the source extraction, the total uncertainty on the 150\,MHz flux density measurements is of order $\sim15$\,per\,cent which is the error shown in the following figures. 

\subsection{The source catalogue}

Source extraction was done using the LOFAR source extraction package PyBDSM (\cite{pybdsm}). This resulted in a catalogue of 5859 sources above a flux limit of 0.75\,mJy. In order to study the spectral index properties of low-frequency radio sources, we first crossmatched the LOFAR catalogue with the deep 1.4 GHz Westerbork mosaic which is closely matched in resolution to the LOFAR data (approximately 15 arcsec). This resulted in a sample of 1366 sources covering the 7 sq. degree field of the Westerbork mosaic. Since the 1.4 GHz dataset is significantly deeper than the LOFAR dataset (rms of 11\,$\mu$Jy/bm), all of the LOFAR sources have a counterpart in the Westerbork 1.4\,GHz catalogue. The median spectral index between 150\,MHz and 1.4\,GHz is $\alpha=-0.8$ (where $S_{\nu}\propto\nu^{\alpha}$), as expected for a low-frequency selected sample (\cite{intema, williams}). 

This sample was then crossmatched with other radio catalogues available in the field: the 15\,GHz 10C survey, 610\,MHz GMRT observations, 345\,MHz Westerbork observations and the 60\,MHz LOFAR image. Since the resolutions of these surveys differ significantly (ranging from 5\,arcsec resolution for the GMRT data, 30 arcsec for the 10C survey and 90 arcsec for the 345\,MHz Westerbork pointing) only point sources were included in this analysis. To ensure we had a reliable sample (and limit the number of spurious detections of peaked sources due to a single incorrect flux measurement) we selected sources that were detected in at least four out of the six frequencies to form our final multi-frequency sample. This sample consists of 117 objects.

\begin{figure*}
\begin{center}
\begin{minipage}{0.48\linewidth}
\includegraphics[width=\linewidth]{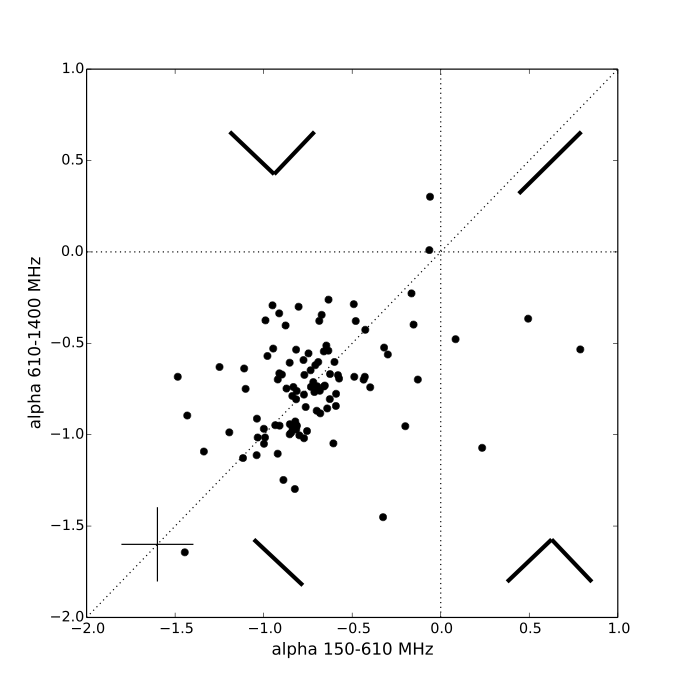}
\end{minipage}
\begin{minipage}{0.48\linewidth}
\includegraphics[width=\linewidth]{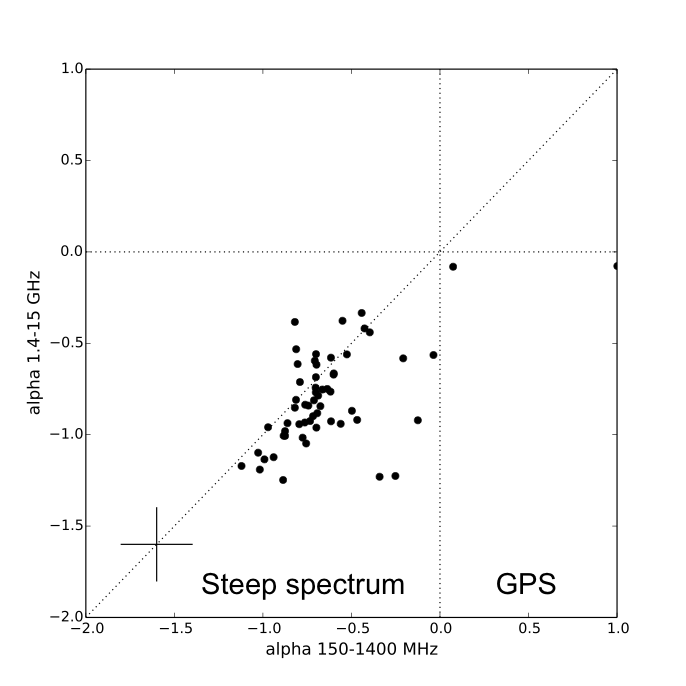}
\end{minipage}
\caption{{\bf Left:} Spectral index between 150 and 610\,MHz plotted against the spectral index between 610\,MHz and 1.4\,GHz (where $S_{\nu}\propto\nu^{\alpha}$). {\bf Right:} Spectral index between 150\,MHz and 1.4\,GHz against spectral index from 1.4 to 15\,GHz. The dashed line denotes a power-law spectrum from the lowest to the highest frequencies plotted. Typical error bars on the spectral indices are shown in the bottom left of each plot. \label{alphaplots}} 
\end{center}
\end{figure*} 

\begin{figure*}
\begin{minipage}{0.33\linewidth}
\includegraphics[width=\linewidth]{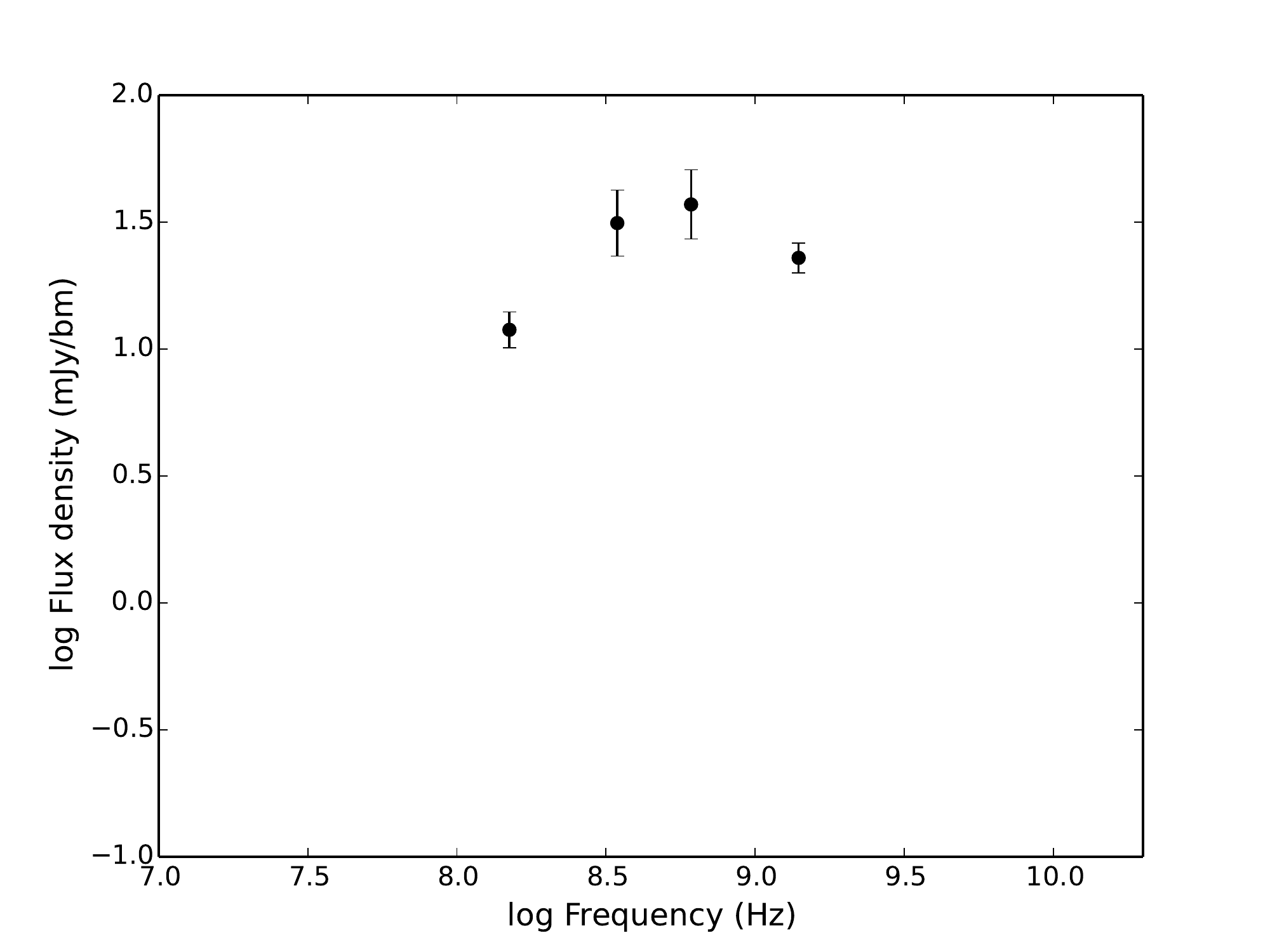}
\end{minipage}
\begin{minipage}{0.33\linewidth}
\includegraphics[width=\linewidth]{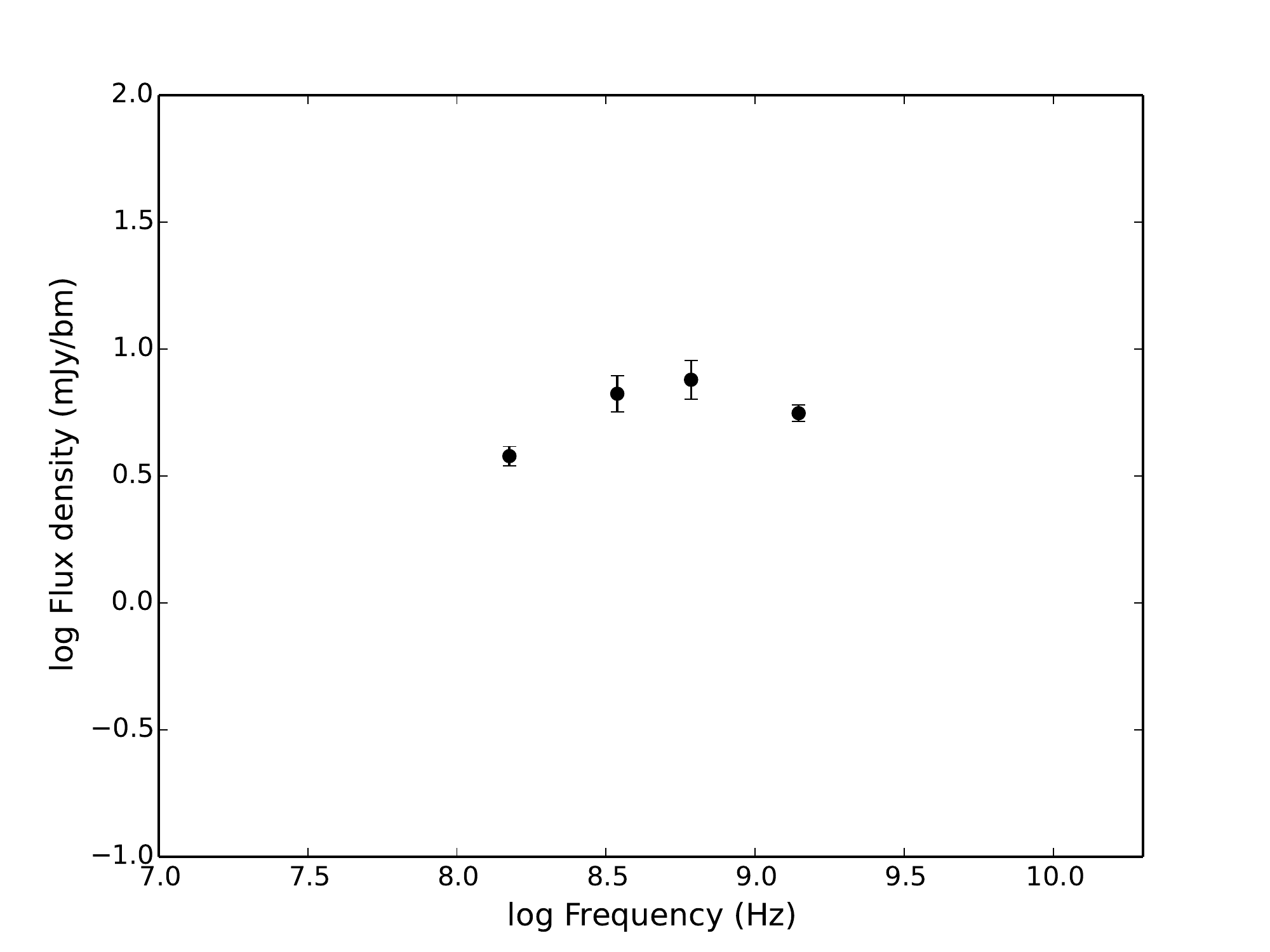}
\end{minipage}
\begin{minipage}{0.33\linewidth}
\includegraphics[width=\linewidth]{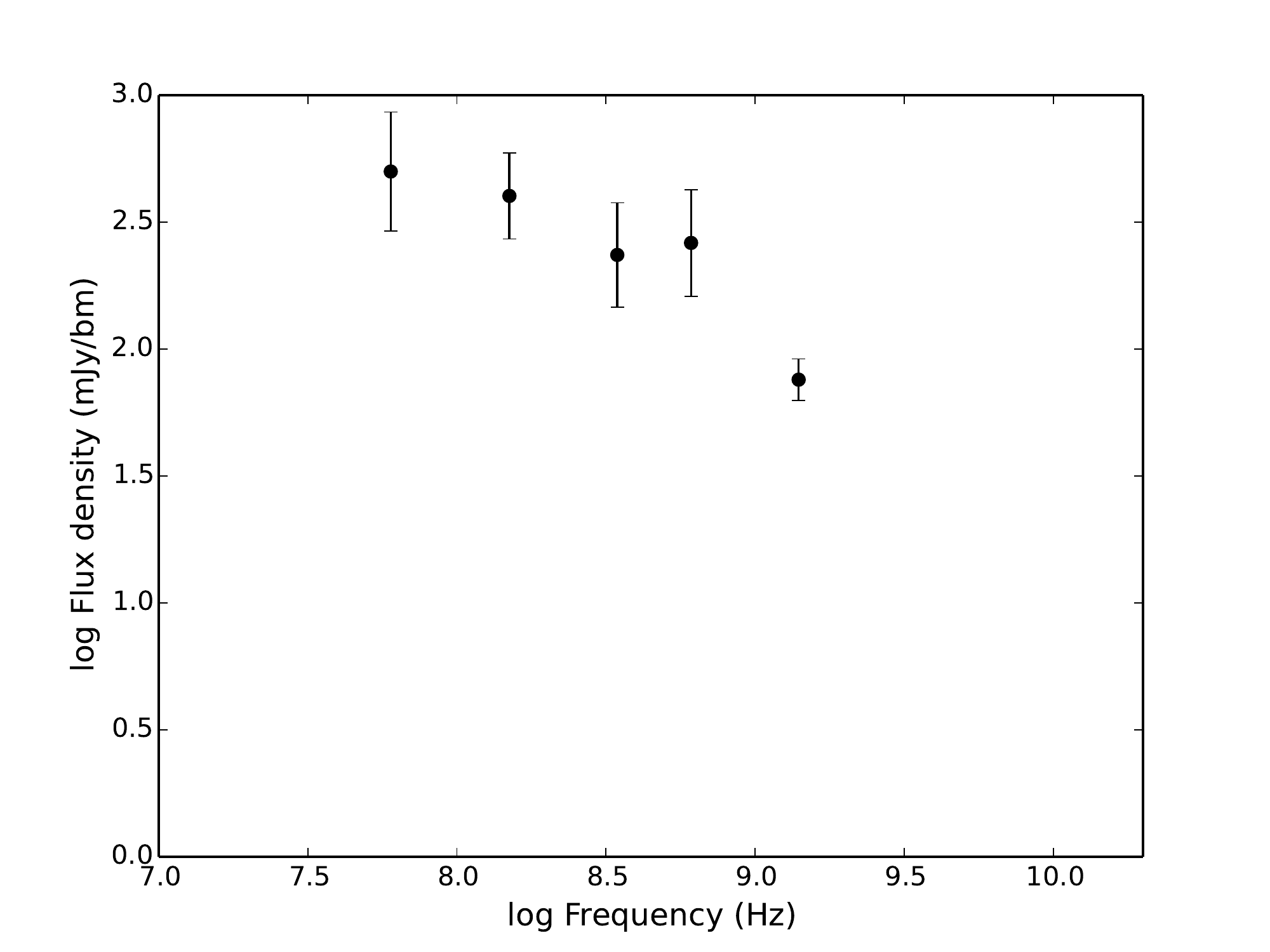}
\end{minipage}
\begin{minipage}{0.33\linewidth}
\includegraphics[width=\linewidth]{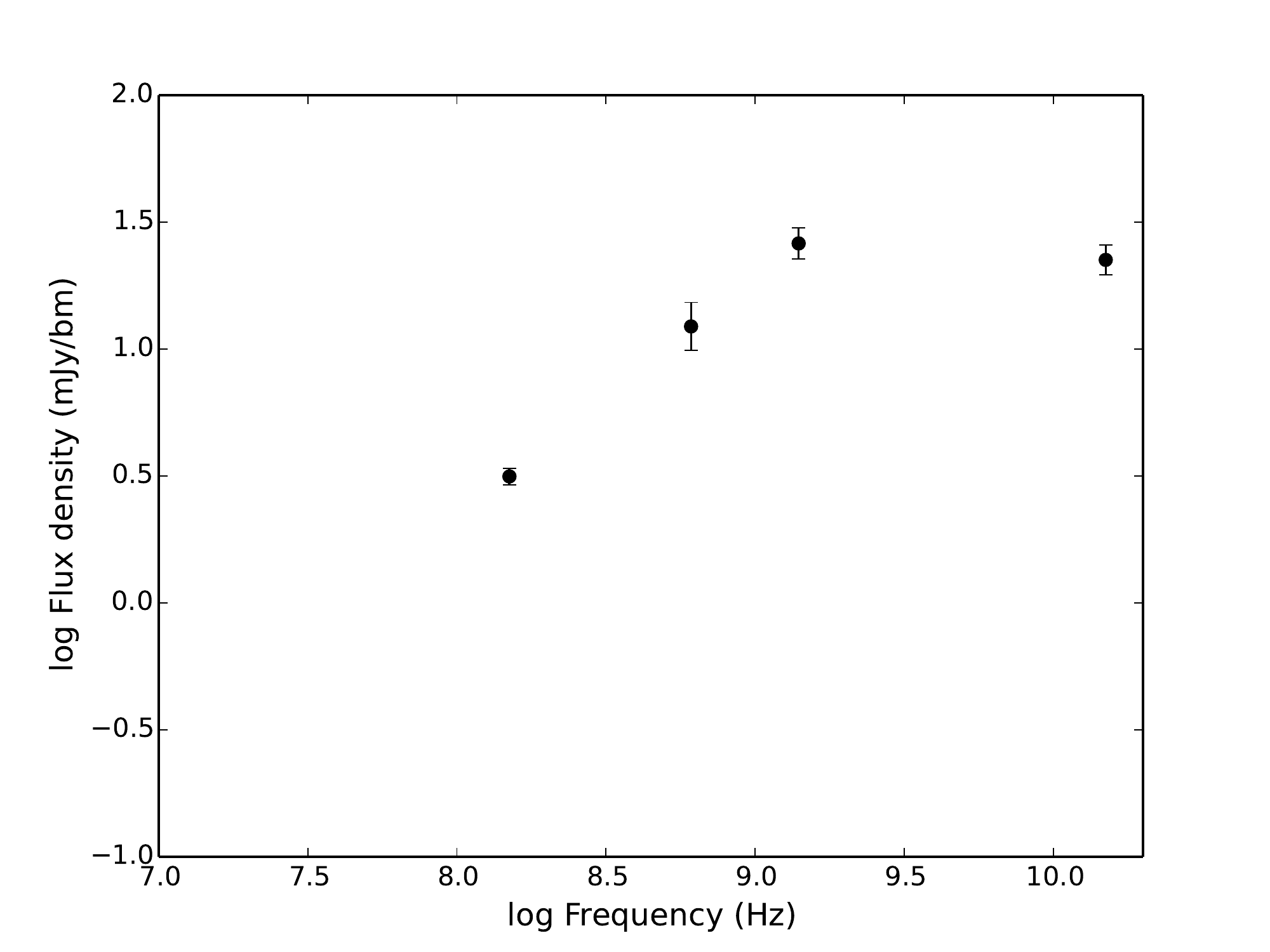}
\end{minipage}
\begin{minipage}{0.33\linewidth}
\includegraphics[width=\linewidth]{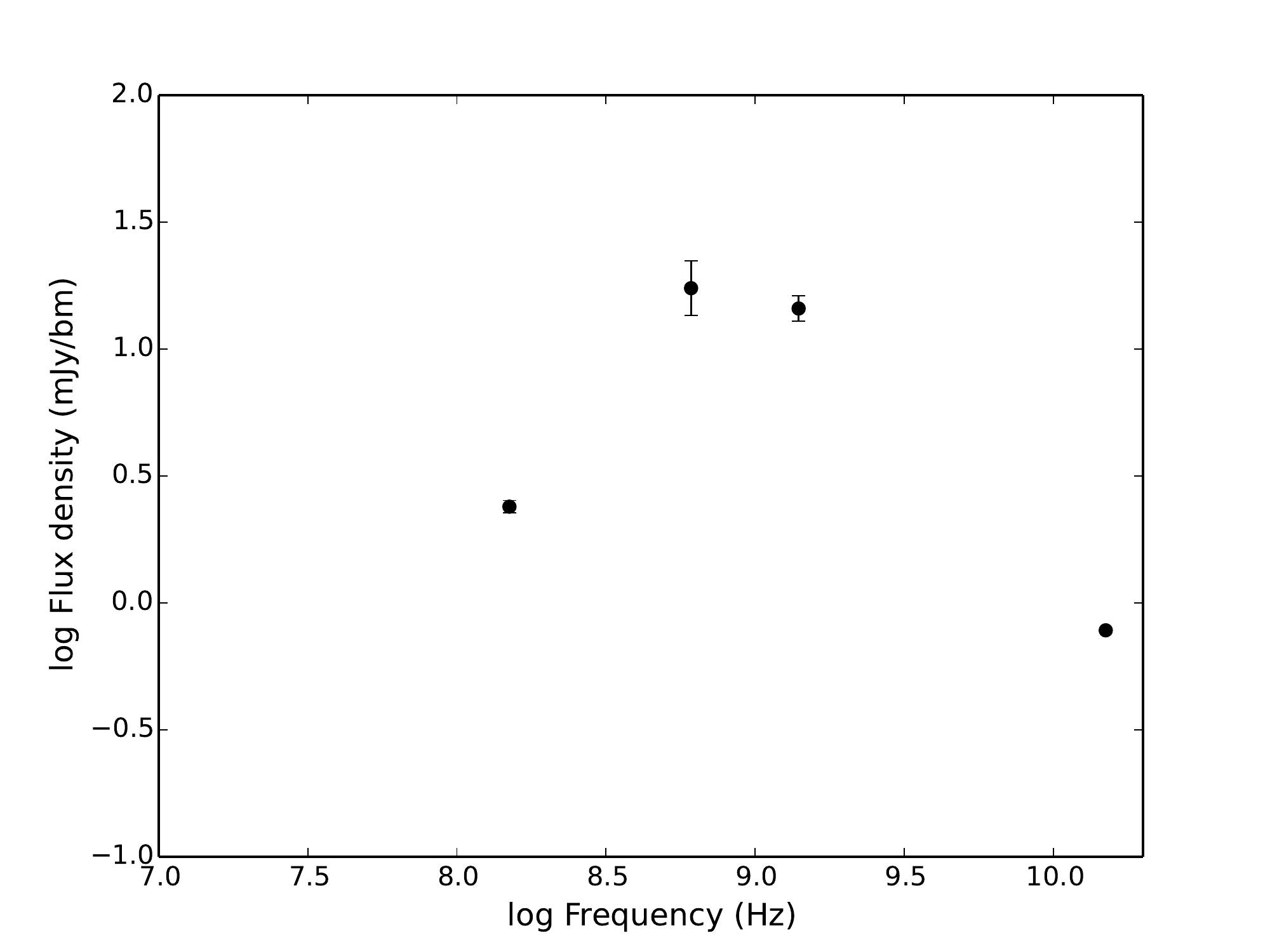}
\end{minipage}
\begin{minipage}{0.33\linewidth}
\includegraphics[width=\linewidth]{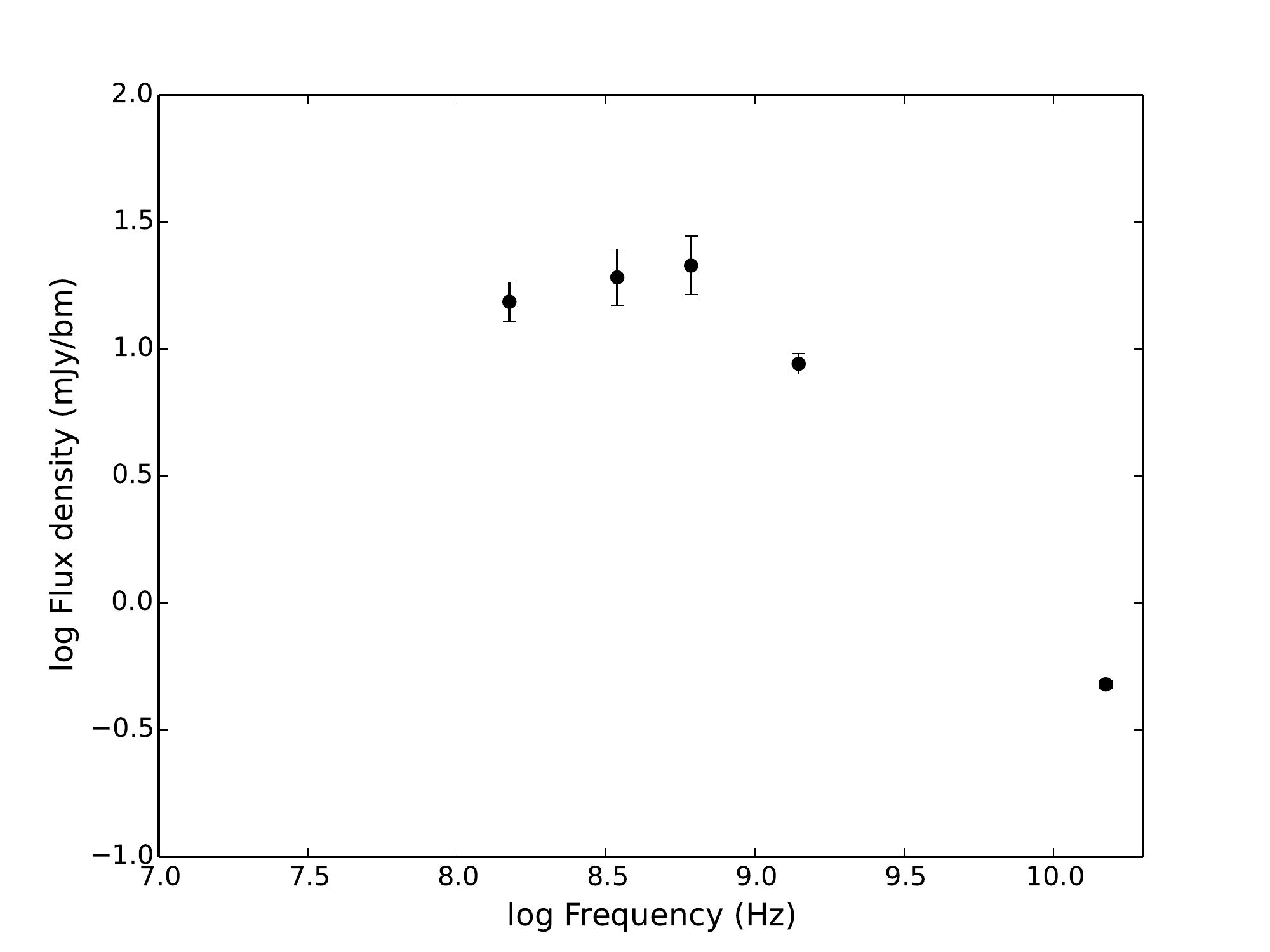}
\end{minipage}
\caption{Radio spectra of GPS/CSS sources detected in the Lockman Hole field. \label{seds}} 
\end{figure*}

\section{GPS and CSS sources in the Lockman Hole field}

To identify the sources with peaked spectra, we calculated the spectral indices between each frequency and plotted the radio colour-colour plots for these 117 sources. Figure \ref{alphaplots} shows the spectral indices between 150 and 610\,MHz against the spectral index between 610\,MHz and 1.4\,GHz on the left, and spectral indices between 150\,MHz and 1.4\,GHz against 1.4 and 15\,GHz on the right. The majority of the sample have steep, power-law spectra (marked by the dashed line) and thus lie in the lower-left quadrant of these plots. Sources with peaked spectra are located in the lower-right quadrant. From these plots we identify six new candidate GPS sources in the Lockman Hole field. 

The radio spectra of these six sources are shown in Figure \ref{seds}, showing that the peak frequencies range from 150\,MHz to 1.4\,GHz. While the low-frequency information is essential in identifying these objects as candidate GPS or CSS sources, ancillary data at other wavelengths is needed to determine the nature of these sources. This is needed to firstly confirm that these are true GPS or CSS sources, rather than variable sources that only appear to have a peaked spectrum, but also to distinguish between high-z GPS sources, or more local CSS sources. The literature was searched for existing information on these sources, but none of them have known redshifts. Only 1 source is detected in the SDSS survey with a g-band magnitude of 21.6 and another two sources are detected in the Spitzer Extragalactic Representative Volume Survey (SERVS; \cite{servs}) with 3.6\,$\rm\mu$m fluxes of 30 and 5\,$\rm\mu$Jy respectively. Unfortunately we cannot place limits on the non-detected sources as these were outside of both the SERVS and the larger SWIRE footprints. To further investigate these sources, follow-up optical/IR observations are needed to identify the host galaxies and find the redshift. Alternatively, high-resolution VLBI imaging would allow us to place useful limits on the linear size of these objects, providing an approximate age of the radio source.

\begin{figure}
\centering{\includegraphics[width=0.9\linewidth]{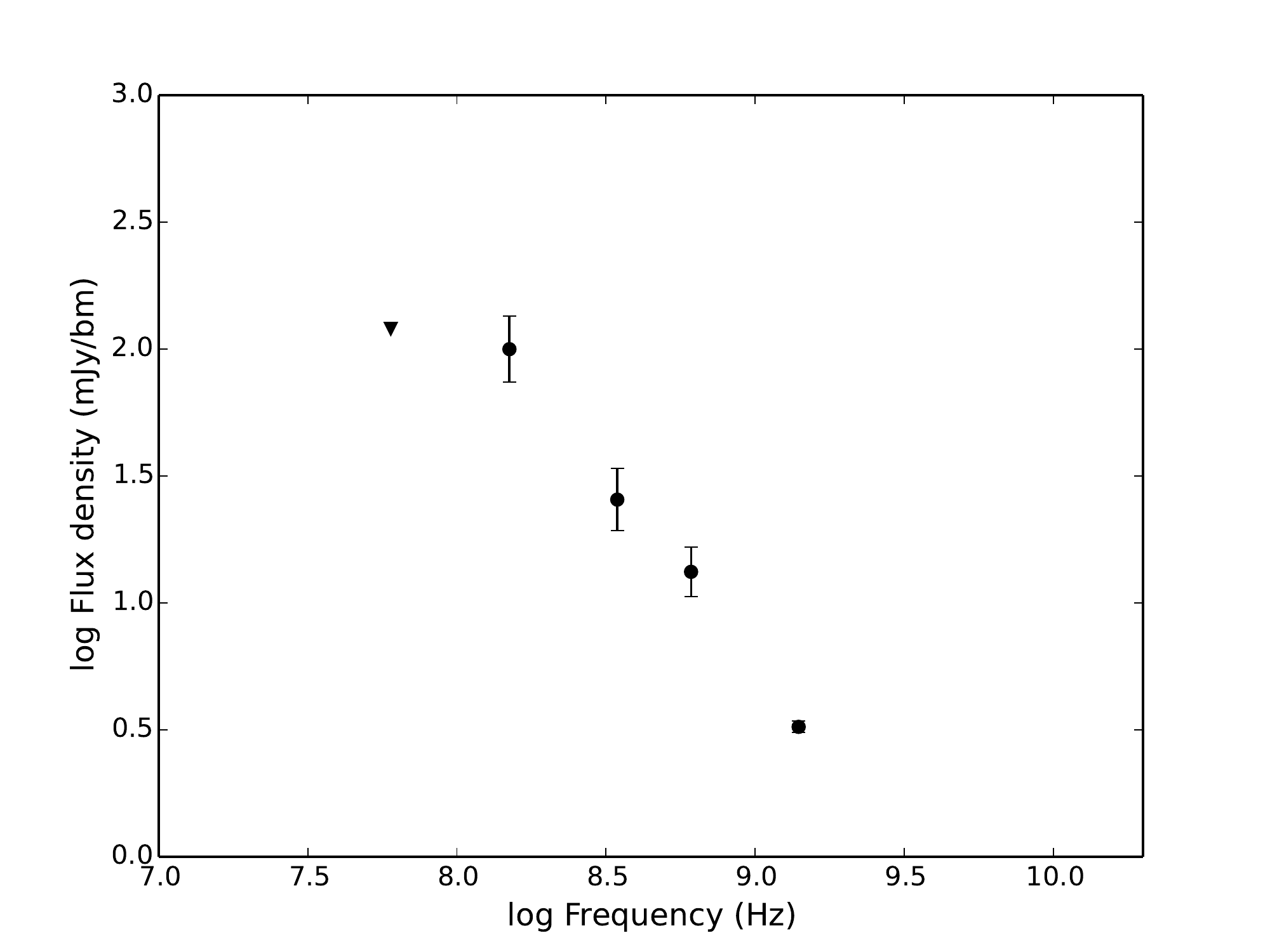}}
\caption{An additional peaked spectrum source not identified in the colour-colour plots due to the non-detection at 60\,MHz. This source exhibits an ultra steep spectrum of $\alpha_{150}^{1400}=-1.6$ between 150 and 1.4\,GHz, but has a much flatter spectral index from 60--150\,MHz of $\alpha_{60}^{150}>-0.2$ . \label{uss}} 
\end{figure}

One of the limitations in using the radio colour-colour plots (shown in Figure \ref{alphaplots}) to identify GPS candidates is that it may miss objects where we only have an upper limit on the spectral peak. To account for this, the radio spectra of each source was visually inspected to search for peaked-spectrum sources that may have been missed in the colour-colour plots. One such example is shown in Figure \ref{uss}. This object is detected at 100\,mJy in the LOFAR HBA image and has an ultra-steep spectrum of $\alpha_{150}^{1400}=-1.6$ between 150\,MHz and 1.4\,GHz. However, it was not detected in the 60\,MHz LBA image. Using a 5 sigma upper limit of 120\,mJy at 60\,MHz we can place a lower limit of the spectral index between 60--150\,MHz of $\alpha_{60}^{150}>-0.2$. This suggests that this particular source turns over quite steeply somewhere between 60 and 150\,MHz. Note that this source is included in the colour-colour plots shown in Figure \ref{alphaplots} (the point with the steepest spectral index in the plot on the left), but it doesn't fall into the peaked-spectrum quandrant as the peak occurs between 60 and 150\,MHz.

\section{Preparing for future surveys with LOFAR}

The data presented here reaches the typical depth required for the wide-area (2$\pi$ steradian) `Tier-1' LOFAR survey. As such, the results obtained from the analysis of the Lockman Hole region can provide insight into how many GPS/CSS sources we can expect to detect in the complete LOFAR Tier-1 survey. Limiting our analysis to the 7 square degree field covered by the 1.4\,GHz Westerbork mosaic, we detect seven MHz-peaked spectrum sources. Extrapolating this to the full northern sky we could optimistically expect to detect as many as 20,000 GPS/CSS sources. However, our ability to detect these sources is not only limited by the flux limit reached in the LOFAR observations, but also by the complementary multi-wavelength data available. The Lockman Hole field was selected precisely because of the deep, multi-frequency radio data available in this region and therefore is not representative of the data available for the full survey. As such, we would probably detect only a fraction of this number in the LOFAR Tier-1 survey.

Nonetheless, this preliminary analysis of sources in the Lockman Hole field provides insight into the population of GPS and CSS sources we can expect to detect with the advent of the new low-frequency telescopes now available such as LOFAR, the Murchison Widefield Array (MWA; \cite{MWA}) and the Long Wavelength Array (LWA; \cite{LWA}). By probing the lowest-frequency radio regime we can trace the spectral peak to much lower frequencies, allowing us to better explore the evolution of these young radio sources.

\acknowledgements

The research leading to these results has received funding from the European Research Council under the European Union's Seventh Framework Programme (FP/2007-2013) / ERC Advanced Grant RADIOLIFE-320745. We also thank the referee for useful suggestions which helped improve these proceedings.

% Use this code if you wish to generate your bibliography with BibTeX;
% please replace first the string "an-demo" below with the name(s) of
% the BibTeX data base(s) you want to use.
% The resulting bibliography-output (the contents of the .bbl file)
% must be pasted into this file before submission.
% 
% \bibliographystyle{an}
% \bibliography{an-demo}

\begin{thebibliography}{}
%  \bibitem{} Author1, A.B., Author2, C.D.: 2001, AN 322, 1
%  \bibitem{} Author3, E.F., Author4, G.H.: 2001, AN 322, 10
%  \bibitem{} Author5, I.: 2001, AN 322, 20
%  \bibitem{} Author6, J.: 2001, AN 322, 30

  \bibitem[AMI consortium, 2011]{10c}AMI consortium 2011, MNRAS 415, 2699
  \bibitem[Bicknell et al., 1997]{bicknell} Bicknell, G. V., Dopita, M. A., O'Dea, C. P., 1997, ApJ, 4
  \bibitem[Brunner et al., 2008]{xmm} Brunner, H., et al., 2008, A\&A 479, 283	
  \bibitem[Coppejans et al., 2015]{coppejans} Coppejans, R., Cseh, D., Williams, W. L., van Velzen, S., Falcke, H., 2015, MNRAS 450, 1477 
  \bibitem[de Vries et al., 2008]{devries} de Vries, N., Snellen, I. A. G., Schilizzi, R. T., Mack, K.-H., Kaiser, C. R., 2009, A\&A 498, 641
  \bibitem[Ellingson et al., 2013]{LWA} Ellingson, S. W., et al., 2013, IEEE Transactions on Antennas and Propagation 61, 2540
  \bibitem[Falcke, K¨ording \& Nagar 2004]{falcke}Falcke H., K¨ording E., Nagar N. M., 2004, New Astron. Rev. 48, 1157
%  \bibitem[Fanti et al., 1990]{fanti} Fanti, R., Fanti, C., Schilizzi, R. T., Spencer, R. E., Rendong, N., Parma, P., van Breugel, W. J. M., Venturi, T., 1990, A\&A 231, 333
  \bibitem[Fanti et al., 1990]{fanti} Fanti, R., et al., 1990, A\&A 231, 333
  \bibitem[Fotopoulou et al., 2012]{Fotopoulou} Fotopoulou, S., et al., 2012, ApJS 198, 1
  \bibitem[Garn et al., 2010]{garn} Garn, T. S., Green, D. A., Riley, J. M., Alexander, P, 2010, BASI 38, 103 
  \bibitem[Guglielmino et al., 2012]{2012rsri.confE..22G} Guglielmino G., Prandoni I., Morganti R., Heald G., 2012, rsri.conf, 22 
  \bibitem[Guglielmino et al., 2014]{guglielmino} Guglielmino, G., Prandoni, I., Morganti, R., Heald, G., Mahony, E., van Bemmel, I., 2014, IAU Symposium 304, 108
  \bibitem[Heald et al., 2010]{pipeline} Heald, G., et al.: 2010, arXiv:1008.4693
  \bibitem[Intema et al., 2011]{intema} Intema, H.~T., van Weeren, R.~J., R{\"o}ttgering, H.~J.~A., Lal, D.~V., 2011, A\&A 535, 38 
  \bibitem[Lockman, Jahoda \& McCammon, 1986]{lockman} Lockman, F. J., Jahoda, K., McCammon, D., 1986, ApJ 302, 432
  \bibitem[Mauduit et al., 2012]{servs} Mauduit, J. C. et al., 2012, PASP 124, 714 
  \bibitem[Mohan \& Rafferty, 2015]{pybdsm} Mohan, N., Rafferty, D., 2015, Astrophysics Source Code Library 1502.007
  \bibitem[O'Dea 1998]{odea} O'Dea C. P., 1998, PASP, 110, 493
  \bibitem[Offringa et al., 2012]{offringa} Offringa, A. R., van de Gronde, J. J., Roerdink, J. B. T. M., 2012, A\&A 539, 95
  \bibitem[Oliver et al., 2012]{hermes} Oliver, S. J., et al,, 2012, MNRAS 424, 1614 
%  \bibitem[Pandey et al., 2009]{bbs} Pandey, V. N., van Zwieten, J. E., de Bruyn, A. G., Nijboer, R, 2009 in "The Low-Frequency Radio Universe", ASP Conference Series 407, 384
  \bibitem[Pandey et al., 2009]{bbs} Pandey, V. N., van Zwieten, J. E., de Bruyn, A. G., Nijboer, R, 2009, ASP Conference Series 407, 384
  \bibitem[Polletta et al., 2006]{chandra} Polletta, M., et al., 2006, ApJ 642, 673
  \bibitem[Roger et al., 2013]{rcb} Roger, R. S., Costain, C. H., Bridle, A. H., 1973, AJ 78, 1030
  \bibitem[Scaife \& Heald, (2012)]{scaife} Scaife, A.~M.~M., Heald, G.~H., 2012, MNRAS Letters 423, L30 
  \bibitem[Snellen et al. 2000]{snellen} Snellen I. A. G., Schilizzi R. T., Miley G. K., de Bruyn A. G., Bremer M. N., R¨ottgering H. J. A., 2000, MNRAS, 319, 445
  \bibitem[Tasse et al., 2013]{awimager} Tasse, C., van der Tol, S., van Zwieten, J., van Diepen, G., Bhatnagar, S., 2013, A\&A 553, 13
  \bibitem[Tingay et al., 2013]{MWA} Tingay, S. J., et al., 2013, PASA 30, 1
  \bibitem[Tingay et al., 2015]{tingay} Tingay, S. J., et al., 2015, AJ 149, 74
  \bibitem[Van Haarlem et al., 2013]{lofar} Van Haarlem, M. P., et al., 2013, A\&A 556, 2
  \bibitem[Whittam et al., 2013]{whittam} Whittam I.~H., et al., 2013, MNRAS 429, 2080 
  \bibitem[Williams et al., 2013]{williams} Williams, W.~L., Intema, H.~T., R{\"o}ttgering, H.~J.~A., 2013, A\&A 549, 55

%  \bibitem[\protect\citeauthoryear{Guglielmino et al.}{2014}]{2014IAUS..304..108G} Guglielmino G., Prandoni I., Morganti R., Heald G., Mahony E., van Bemmel I., 2014, IAUS, 304, 108 
\end{thebibliography}
% 
% Replace the following example bibliography with your references
% before submission:

\end{document}